\documentclass[prb,11pt]{revtex4-1}

\usepackage{amssymb}  %
\usepackage{amsthm}
\usepackage{amsmath}    
\usepackage{graphicx}   
\usepackage{epstopdf}  
\usepackage{verbatim}   
\usepackage{hyperref}   
\newtheorem{theorem}{Theorem}
\theoremstyle{definition}
\newtheorem{definition}{Definition}

\begin{document}

\title{A Conic Section Approach to the Relativistic Reflection Law }
\author{Mohsen Maesumi (\url{maesumi@gmail.com})}
\affiliation{Lamar University, Beaumont, Texas 77710, USA}

\date{\today}

\begin{abstract}
We consider the reflection of light, from a stationary source, off of  a uniformly moving flat mirror, and derive the relativistic reflection law using well-known properties of conic sections. The effective surface of reflection (ESR) is defined as the loci of intersection of all beams, emanating from the source at a given time, with the moving mirror. Fermat principle of least time is then applied to ESR and it is shown that, assuming the independence of speed of light,  the result is identical with the relativistic reflection law. For a uniformly moving mirror ESR is a conic and the reflection law becomes a case of bi-angular equation of the conic, with the incident and reflected beams coinciding with the focal rays of the conic. A short calculus-based proof for accelerating mirrors is also given.
\end{abstract}

\maketitle
\section{Introduction}

The study of the reflection of light from a moving mirror has been a fundamental theme since the inception of relativity. In particular the reflection law  was obtained using  the Lorentz transformation \cite[p. 915]{Einstein} as 
\begin{equation}\label{eq: Ein}
\cos\phi''' = - {{(1+({v\over V})^2)\cos \phi  -2 {v \over V}} \over {1 -2 {{v}\over{V}}\cos \phi +{({v\over V})^2}}} ,
\end{equation}
where $V$ is the speed of light in vacuum, $v$ is velocity of mirror, $\phi$ is the angle of incidence and $\phi'''$ is the angle of reflection. The angles and the mirror velocity are measured with respect to the same axis normal to the mirror.

There are derivations of  the reflection  law without using the Lorentz transformation \cite{Gj}, however here  we focus  on a geometric approach using well-known properties of conic sections.   

Consider  a ray of light that is emanated from a point source at position $F$ at time $\tau$  and reaches a second point $G$ after reflection from the moving  mirror. Fermat principle requires the trajectory to be such that it minimizes the travel time over all paths. These paths start at $F$ at $\tau$ and bend at the point of intersection with the moving mirror before reaching $G$.
This prompts the following definition: 

\begin{definition}[ESR]
The effective surface of reflection, $ \rm{ESR}(\tau)$,  is the  loci of the intersection of all beams, emanating from the stationary source at a given time $\tau$,  with the moving mirror. 
\end{definition}
We propose to derive the relativistic reflection law by applying  Fermat principle  to ESR. 

\begin{definition}[Fermat Principle]
  If a ray of light is emanated from a point source at a position $F$ at time $\tau$  and reaches a second point $G$ after reflection from a mirror then the path it takes is the one minimizing the time from $F$ to  $\rm{ESR}(\tau)$ to $G$. 
\end{definition}

To measure the time of flight we need the speed of light before and after the incidence.
The second principle of relativity states that the speed of light in vacuum  is independent of the movement of its source.
In particular the incident and reflected beam travel at the same speed in vacuum. Now we are ready to state the main result of this paper.

\begin{theorem}
If the speed of the reflected beam is same as the incident beam then the reflection angle calculated by the  Fermat principle is same as the one predicted by the special theory of relativity.
\end{theorem}

ESR, for a  flat mirror moving at a constant velocity $v$,  is a conic.  The location of the mirror at the time of emanation is the linear directrix, the source is a focus, and the ratio of speed of light to the speed of the mirror is the eccentricity. 
Hence the incident and the reflected rays coincide with the focal rays of the conic. 
The reflection law then turns out, in fact, to be the bi-angular equation  \cite{Bi} of the conic.
We offer two proofs, one using basic geometry, for a uniformly moving mirror, and another, using calculus, for accelerating mirrors. The  Lorentz  transformation is not used. 

The solution presented is applicable to any signal bouncing from a moving boundary so long as the reflected signal has the same speed as the incident signal. While this is standard in relativity it may also occur in non-relativistic classical  cases. The final section of paper discusses some of  the physics aspects of the problem.
We assume  the source resides in vacuum or an isotropic stationary medium. In the latter case we consider a thought experiment  where the movement of the mirror does not cause any motion or turbulence  in the medium. The speed of light in the medium in any direction and from any source is the same constant.

\section{Reflection from a  Uniformly Moving Mirror}
Consider a stationary source of light at $F=(0,0)$,  and a mirror, parallel to the $y$-axis, that moves in the direction of $x$-axis with a uniform velocity of $v$ and at time $t$ is located at $x(t)=v t+x_0$. At time $\tau$ the source emits a ray of light at an angle of $i$.  We find $r$, the angle of reflection  from the mirror using Fermat principle and compare it with \eqref{eq: Ein}.

 The comparison of the speed of light (or signal) in the medium, $c$, with the speed of the mirror $|v|$ results in three cases: a conventional hyperbolic case when $|v|<c$, a hypothetical elliptic case when $|v|>c$, and a threshold parabolic case when $|v|=c$. 

\subsection{The Hyperbolic Case}
\begin{figure}
\centering
\includegraphics[width=\textwidth]{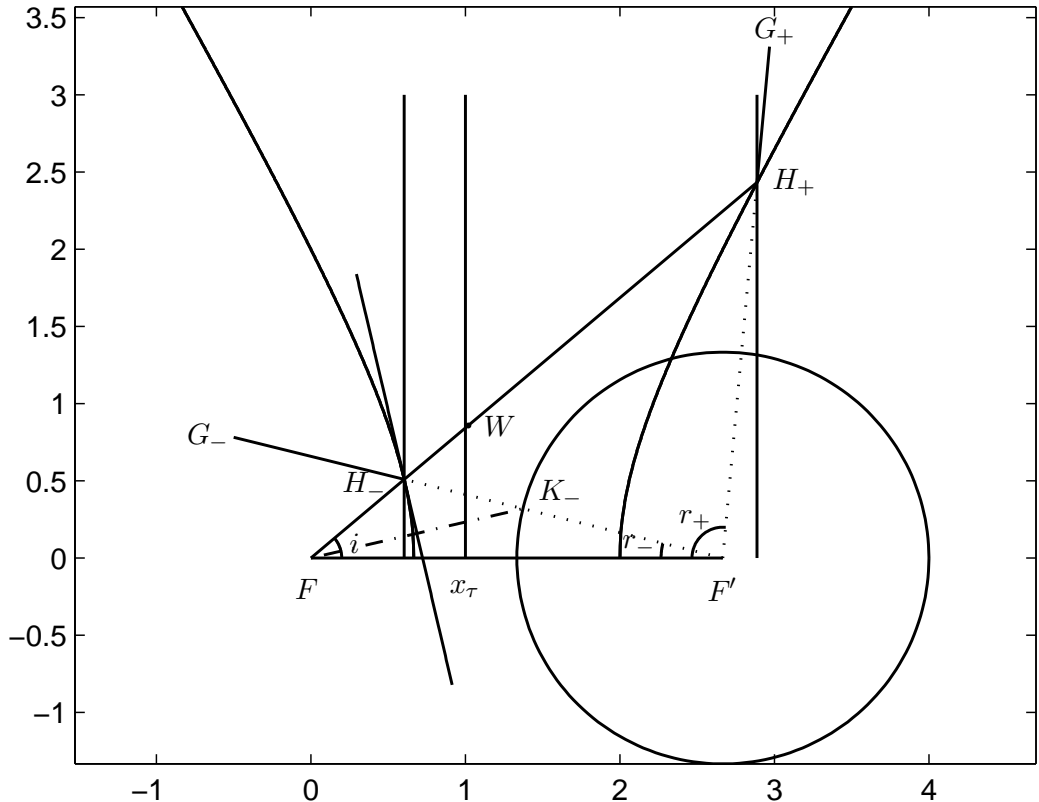}

\caption {The two branches of a semi-hyperbola, the one on the  left corresponds to an initially approaching mirror with $\beta= -0.5$ and the one on the right corresponds to a receding mirror with $\beta=+0.5$. The location of mirror at the emanation of light  is  $x=x_\tau$. Light emanates from $F$ and for an incidence angle of  $i=0.7$ the points of incidence $H_-$ and $H_+$  with the mirrors are depicted. The tangent at $H_-$ and the reflection of $F$ in the tangent, namely point $K_-$ on the right circular directrix, are drawn.
Fermat principle appears as $G_- H_- K_- F'$ being a straight line.  
 All incident and reflected rays travel to the left of the mirror even when $i$ or $r_+$ are obtuse.
} 
\label{fig:mirrorh1.eps}
\end{figure}

In this case the speed of light in the given medium is  more than that of the mirror, $c >|v|>0$. See Figure ~\ref{fig:mirrorh1.eps}. Let the incident beam strike the mirror at $H$ ($H_+$ when $v>0$ and $H_-$ when $v<0$). We show that $H$ resides on a branch of a certain hyperbola (depending on $x(\tau)$ and $v/c$)
with $F$ as a focus; and identify the secondary focus $F'$.  The incident beam $FH$ will reflect so that the reflection  coincides with the extension of  the secondary focal ray $F'H$. As if the light reflected from the hyperbola or that it emanated from an image at $F'$. 

Let the particle of light leave the source at $\tau$, when the mirror is at $x_\tau=v\tau+x_0$, and strike the mirror at time $T$, at the point $H=(x,y)$. Show the length of $FH$ by $|FH|=\ell$. Then $\ell =c(T-\tau)$ and $x-x_\tau=v(T-\tau)$.
Therefore $(x-x_\tau)/\ell=v/c$.  Let $\beta = v/c$ and note that $0<|\beta|<1$, then the loci of such points $H$, for all incident angles $i$, is a branch of  a hyperbola.

Specifically,  let $\alpha=\sqrt{1-\beta^2}$, $a=|\beta x_\tau|/\alpha^2$, $b=|x_\tau|/\alpha$, $d=x_\tau/\alpha^2$, then the equation of the hyperbola is $(x-d)^2/a^2-y^2/b^2=1$ with center $(d,0)$,
 asymptotes $y=\pm (\alpha  / \beta)(x-d) $ 
at an angle of $\arccos(\pm \beta)$,  
vertices   $(x_\tau/(1\pm \beta),0)$,
 focal length $\bar c=\sqrt{a^2+b^2}=|x_\tau|/\alpha^2$, 
eccentricity $e=\bar c/a=1/|\beta|$, 
linear directrix $x=x_\tau$,  
source focus $F=(0,0)$, 
and image focus $F'=(2 x_\tau/\alpha^2, 0)$.

(In what follows we assume the left side of  the mirror is reflective, and require $x_\tau>0$. The  right branch of the  hyperbola then corresponds to a receding mirror, $v>0$, and the left  branch corresponds to an initially approaching mirror, $v<0$. The incidence angle $i$ is the angle of $FH$ with the focal axis $FF'$ or the $x$ axis.  In this section the reflection angle $r$ is  defined as  the angle of $F'H$ with $F'F$, or $-x$, to conform with the pedagogical convention, where, for a stationary mirror, $r=i$. In  \cite{Einstein},  both angles are measured with respect to the $x$ axis, and $\phi'''$ in that derivation is equal to  $\pi -r$ in this section. Hence $\cos \phi'''=-\cos r$.)

 The reflection law  is a description of the hyperbola in terms of $i$ and $r$ (and a parameter such as the eccentricity).
One way to obtain this law is to solve the triangle $F'FH$ by following  the steps of its geometric construction. Assume we are on the right branch. We know  one side, $|FF'|=2\bar c$, the difference of the other sides,  $|FH|-|F'H| =2a$, and the angle $ \measuredangle F'FH =i$ . The construction is possible if $\cos i>a/ \bar c$. 
 To construct the triangle draw the ray $FW$ with $ \measuredangle F'FW =i$ and
 $|FW|=2a$. To find $H$  build the isosceles triangle $F'WH$ with base $F'W$ and side $WH$ as the extension of $FW$.

To calculate the  reflection angle $ r=\measuredangle FF'H $ let $j= \measuredangle FWF'$, $k= \measuredangle WF'F$, and $2w=|F'W|$. Apply laws of cosines and sines to the triangle $FWF'$ to get $\cos j = (a^2+w^2-\bar c^2)/2aw$, $\cos k = (\bar c^2+w^2-a^2)/2\bar c w$, $\sin j = (\bar c /w)  \sin i$, $\sin k=(a/w) \sin i$. We have $r=\pi-j+k$ therefore $\cos r= -(\cos j \cos k+\sin j \sin k)$. Substituting 
$w^2= a^2+{\bar c}^2-2a{\bar c} \cos i$
and simplifying gives the relativistic reflection formula

\begin{equation}\label{eq: FerC}
\cos r= \frac{(\beta ^2 +1) \cos i - 2 \beta }{(\beta^2+1) -2 \beta \cos i}.
\end{equation}
The same result applies to the left branch of the hyperbola for which $\beta$ is negative.  This concludes the proof of Theorem 1.

While \eqref{eq: FerC} is the common form of the relativistic reflection law, other forms are possible.  For example,
in the triangle $F'FH$ let $\ell=|FH|$ and $\ell'=|F'H|$, and use the law of sines  
$\ell / \sin r= \ell' / \sin i = 2\bar c / \sin(\pi-i-r)$  to substitute for $\ell$ and $\ell'$ in $\ell -\ell'=2a$. 
This results in 
\begin{equation}\label{eq: FerS1}
\sin r -\sin i = \beta \sin(r+i),
\end{equation}
or equivalently
\begin{equation}\label{eq: FerS2}
\sin (r-i)/2 = \beta \sin (r+i)/2,
\end{equation}
where $(r-i)/2$ is the angle between the normals to the mirror and to the hyperbola, and $(r+i)/ 2$ is the angle between the incident or the reflected ray and the normal to the hyperbola. 

Similarly, if we define the  angle $z\in(0,\pi)$ as the value of $r$ when $i=\pi/2$, i.e. through $\cos z = -2\beta /(1+\beta^2)$,
then (2) reads $\cos r = (\cos i +\cos z)/(1+\cos i \cos z)$.
The latter can be written as $[(1-\cos r)/(1+\cos r)] =[  (1-\cos z)/(1+\cos z)]  [(1-\cos i)/(1+\cos i)]  $ and since
$(1- \cos x) /(1+\cos x) = \tan^2 (x/2)$  
one gets 
\begin{equation}\label{eq: FerT}
\tan(r/2)=\tan(z/2) \tan(i/2).
\end{equation}

\subsection{The Elliptic Case}
Now we consider the case where  the mirror moves faster than light in the given medium, $|v|>c$. (A discussion on physical plausibility appears in the last section.)
We will see that here the reflected  beam of  light is on the same side of the normal to the mirror as the incident beam. Another oddity is that the image is on the same side of the mirror as the source and is real.

\begin{figure}
\centering
\includegraphics[width=\textwidth]{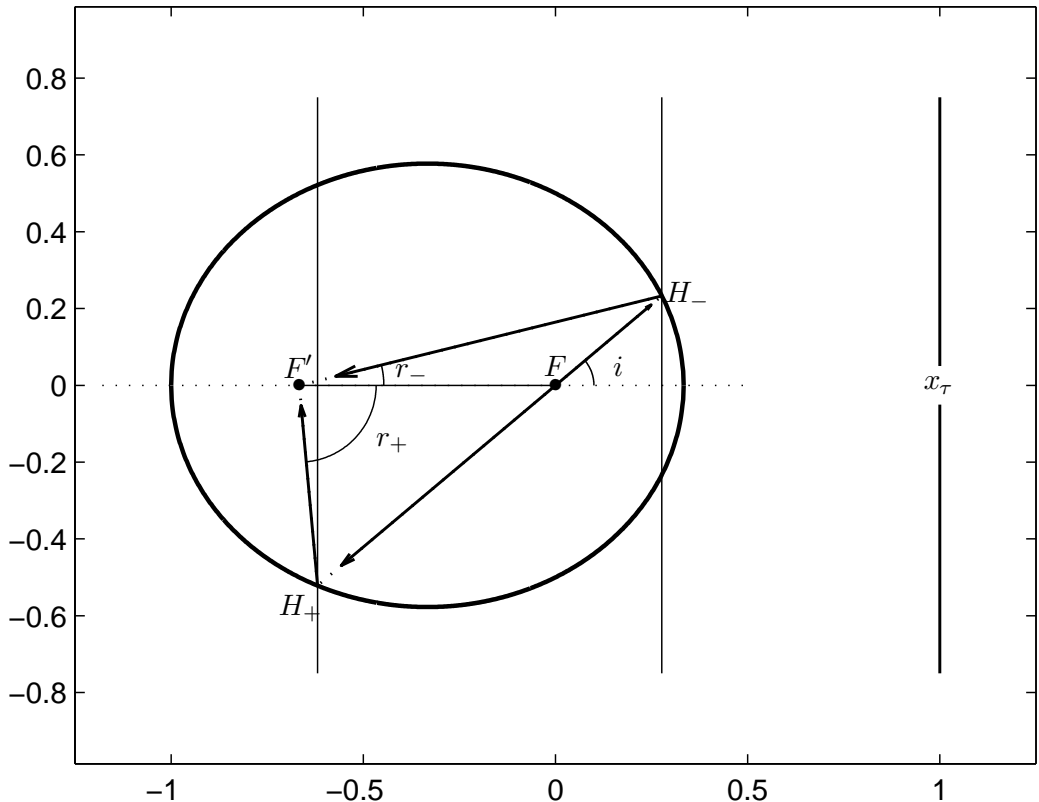}

\caption {The mirror, originally at $x_\tau$,  moves at twice the speed of light, $\beta =-2$. The incident ray $ {FH_-}$, with $i=0.7$, is reflected as $H_- F^\prime$. The reflected beam lags behind the moving mirror.} 
\label{fig:ellipse.eps}
\end{figure}

See Figure ~\ref{fig:ellipse.eps}.
Assume the incident beam strikes the mirror at $H$ (the ray with the angle $i$ at $H_ -$ and the ray with the  angle $\pi-i$ at $H_+$), we show that $H$ resides on a certain ellipse
with $F$ as a focus, and identify the secondary focus $F'$.  The incident beam $FH$ will reflects  from the mirror along    the secondary focal ray $HF'$. As  if the light reflected from the ellipse.  

At the moment of emanation of light  the mirror should be moving toward the source (otherwise light can not catch up to the  mirror).
Henceforth we assume  $x_\tau>0$ and $v<0$.   
Using the same set up as in the hyperbolic case we arrive at  $(x-x_\tau)/\ell=v/c=\beta <-1 $. The loci of such points $H$, for all incident angles $i$, is an ellipse.

Specifically, let $\alpha=\sqrt{\beta^2-1}$, $a=-\beta x_\tau/\alpha^2$, $b=x_\tau/\alpha$, $d=-x_\tau/\alpha^2$, then the equation of the ellipse is $(x-d)^2/a^2+y^2/b^2=1$ with center $(d,0)$,  major axis  vertices at  $(x_\tau/(1\pm \beta),0)$,
 focal length $\bar c=\sqrt{a^2-b^2}=x_\tau/\alpha^2$, eccentricity $e=\bar c/a=-1/\beta$, 
right linear directrix  $x=x_\tau$,
right or source focus $F=(0,0)$, and the left or image focus $F'=(-2x_\tau/\alpha^2,0)$.

Here we define $i$  as angle of  $FH$   
with the $x$ axis   and $r$ as the angle of  $HF'$ with $-x$, or equivalently the angle of  $F'H$ 
with $x$. To employ  a different approach for the derivation of the reflection law we may write the law of cosines for the triangle $FHF'$.  
Let $\ell=|FH|$ and $\ell'=|F'H|$.
We have $\ell^2=\ell'^2+4\bar c^2-4\bar c \ell' \cos r$ and $\ell'^2=\ell^2+4\bar c^2 +4\bar c \ell \cos i$.
Subtract these to get $2(\ell^2-\ell'^2)=4\bar c (\ell \cos i - \ell' \cos r)$, then
use the ellipse definition, $\ell + \ell'=2a$,
to obtain $4a (\ell - \ell') = -4\bar c(\ell \cos i + \ell' \cos r)$. Now, solve a linear system consisting of the latter equation and  $\ell + \ell'=2a$
to produce
$\ell=2a(a-\bar c \cos r )/(2a+\bar c \cos i -\bar c \cos r)$ and
$\ell'=2a(a+\bar c \cos i )/(2a+\bar c \cos i -\bar c \cos r)$.
 Substituting these expressions in $|F'F|= 2\bar c = \ell' \cos r - \ell \cos i$ and re-arranging produces the same relativistic reflection formula \eqref{eq: FerC}, except that now $\beta <-1$.

Note that \eqref{eq: FerC} is invariant under $\beta \to 1/\beta$, therefore there is a correspondence between the left branch of the hyperbola and the ellipse. This shows an interesting property; for a hyperbola and an ellipse that have reciprocal eccentricities the angles $i$ and $r$ match in value. However, note that in the hyperbolic case $i$ and $r$ are both interior angles of $FHF'$  while in the elliptic case $i$ is the exterior and $r$ is the interior  angle.

\subsection{The Parabolic Case} In this case the speed of mirror matches the speed of light in the given medium, $|v|=c$.
Let $x_\tau>0$ and $v=-c$ then $\beta=-1$ and
 $H$ resides on the parabola $2x_\tau x=x_\tau^2-y^2$ with vertex $(x_\tau/2,0)$ and focus $F=(0,0)$. The reflection law is same as before and simplifies to $r=0$. The reflected beam simply moves horizontally toward left  in tandem with the mirror.

\section{Reflection From An  Accelerating Mirror} In this section we will derive \eqref{eq: Ein} for a mirror in an arbitrary rectilinear motion using calculus. We will see that so long as $\beta$ is defined locally, as the fractional velocity of the  mirror at the time of  light impact, our conclusion remains valid.

Suppose the location of the mirror at time $t$ is $x=f(t)$. Let $\tau$ be the time of emanation of light, $T$ time of impact, $(x,y)$  point of impact, and $\ell=\sqrt{x^2+y^2}$. Then $T=\tau  +\ell / c$ and the loci of intersection of light with the mirror is given implicitly  through

\begin{equation}\label{eq: ESR}
x=f(\tau+  \frac{\sqrt{x^2+y^2}}{c} ).
\end{equation}
This is the effective curve of reflection. Applying Fermat's principle to this curve will produce the reflection formula.

Differentiate the loci curve \eqref{eq: ESR} with respect to $x$ to get $1=f^\prime (T) (x+yy^\prime)/(\ell c)$. Solve this for $y^\prime$ and use $f'(T)/c=\beta$, $x=\ell \cos i$, $y=\ell \sin i$ to get 
$y^\prime =(1/\beta - \cos i)/{\sin i}$, 
where $i$ is the angle of incidence.  Now use the fact that the reflection of a vector $B$ in $A$ is given by $C=2 A ( B\cdot A)/(A\cdot A) -B$. Find the reflection of the incident beam $B=(\cos i, \sin i)$ in the tangent to the loci curve $A=(1,y^\prime)$ to get the reflected beam $C=(\cos r,  \sin r)$. This results in

\begin{align}
\cos r &=   \frac{2\beta - (\beta ^2 +1) \cos i  }{(\beta^2+1) -2 \beta \cos i}, \label{eq: AccC}       \\
\sin r &=  \frac{(1-\beta ^2 ) \sin i  }{(\beta^2+1) -2 \beta \cos i}\label{eq: AccS} .
\end{align}
Note that \eqref{eq: AccC} is identical to \eqref{eq: Ein}. A derivation of \eqref{eq: AccS} using relativistic velocity addition formula appears in \cite[p. 7, 57]{Si}.
 (Here we measured all angles with respect to the $x$ axis, or else we should set $C=(-\cos r, \sin r)$ to follow the same convention we used for \eqref{eq: FerC}.) 

\section{The Physical aspects of the reflection problem}

\subsection{Obtuse angle of incidence or reflection}
 In the stationary reflection problem with $r=i$ (in the pedagogical notation) both angles are limited to $[0,\pi/2]$.   
In the moving version one of the two angles can become obtuse. One advantage of geometrical description is that particular events and range of angles are readily suggested by the figures. For example we see in Fig ~\ref{fig:mirrorh1.eps}  that, for the case of mirror initially moving toward the source, $-1<\beta<0$, as the incident ray rises from $0$ to become parallel to the left asymptote at the obtuse angle $i= i_{c_-}=\arccos(\beta)$ the reflected angle rises from $0$ to the acute angle $r=\arccos(-\beta)$. The obtuse angles of incidence corresponds to $H_{-}$ being to the left of $F$.
 If $i \ge i_{c_ -}$ no reflection will occur.
 (To visualize an obtuse angle of incidence consider a ball crossing the path of a bus and moving in the same general direction as the bus; when the front fender of the bus strikes the ball.)

Similarly, in the case of a receding mirror, $0<\beta <1$, as the incidence angle increases from $0$ to become parallel to the right asymptote at the acute angle  $i= i_{c_+} = \arccos(\beta)$ the reflected angle rises from $0$ to the obtuse angle    $r=\arccos(-\beta)$. An obtuse angle of reflection occurs         if $H_+$ is on the right of $F'$.  If  $i \ge i_{c_+}$ no reflection will occur. 
 (To visualize an obtuse angle of reflection consider a ball rolling toward the back of a moving bus  when it hits  the back fender  and continues to roll toward the bus.)
 
\subsection{Spurious answers}
The critical angles, $i_{c_+}$ or $i_{c_-}$, are not flagged by \eqref{eq: FerC}. For example as $i$ exceeds $i_{c_+}$  the formula continues to predict a reflection angle, even though no reflection occurs. To explain,  note that \eqref{eq: FerC} remains invariant  under simultaneous transformations $\beta\to -\beta$ , $i \to \pi-i$,  and $r \to \pi-r$.  As $i$ exceeds $i_{c_+}$ for a receding mirror the reflected angle $r$ that is produced by \eqref{eq: FerC} is  the supplementary angle of the reflection angle for the case of an initially approaching mirror and a supplementary angle of incidence. In other words, for a line drawn through $F$ with angle $i$ in $(i_{c_+},\pi)$  both $H_-$ and $H_+$ are on the left branch. Point $H_-$ is on the ray with incidence angle $i$ and point  $H_+$ is on the ray with incidence angle $\pi-i$.

\subsection{Mirror moving faster than light}

 In the elliptic case, assuming a conventional mirror is involved, no regular reflection can occur since now  light is not fast enough to separate from the mirror. However  a hypothetical  substance that functions as a ``see-through reflector''  may provide a model.  To explain, consider the reflection of sound waves from a net (as in a tennis racket) that is  moving faster than the speed of sound in air without creating air disturbance while still allowing reflection of sound waves that hit the net. Now, due to the net's porous nature, the reflected beam  can form  behind the net.
 We speculate that such see-through mirrors can be constructed by electronically switching a layer of atoms inside a certain crystal into a temporary mirror. As a layer is switched on and then off then the next layer is switched on, in effect creating a moving mirror. The movement of such a mirror will not be a prohibitive physical movement, and without a barrier that will prevent the formation of the reflected beam on the opposite side of the mirror. Such a scenario will allow for the speed of mirror to be higher than speed of light in the crystal and for the reflected light to form behind the mirror.

\end{document}